\documentclass[a4paper,fleqn]{cas-sc}

\usepackage[numbers]{natbib}
\usepackage{subcaption}
\usepackage{makecell}

\def\tsc#1{\csdef{#1}{\textsc{\lowercase{#1}}\xspace}}
\tsc{WGM}
\tsc{QE}
\tsc{EP}
\tsc{PMS}
\tsc{BEC}
\tsc{DE}

\begin{document}
\let\WriteBookmarks\relax
\def\floatpagepagefraction{1}
\def\textpagefraction{.001}
\shorttitle{}
\shortauthors{H. Lao et~al.}

\title [mode = title]{Food-Embedded Cold Energy Flows in Decentralised Solar Cold Chains}                      



\author[1]{Hange Lao}
\cormark[1]
\credit{Conceptualization, Methodology, Data curation, Software, Visualization, Writing - Original draft preparation, Writing - review\&editing}

\author[2]{Binjian Nie}
\credit{Data curation, Writing - review}

\author[1]{{Wei He}}
\cormark[2]
\credit{Conceptualization, Methodology, Supervision, Writing - review\&editing}

\affiliation[1]{organization={Department of Engineering, King’s College London, London WC2R 2LS, UK}}
\affiliation[2]{organization={Department of Engineering Science, University of Oxford, Oxford, OX1 3PJ, UK}}

\cortext[cor1]{Corresponding author: hange.lao@kcl.ac.uk}

\cortext[cor2]{Principal corresponding author: wei.4.he@kcl.ac.uk}


\nonumnote{The short version of the paper was presented at ICAE2025, Bangkok, Thailand, Dec. 8-12, 2025. This paper is a substantial extension of the short version of the conference paper.}

\begin{abstract}
Reliable cold storage is needed to reduce meat loss in informal food systems, but conventional cold-chain expansion is difficult where electricity supply is weak and battery-based solar refrigeration is costly. This study develops an hourly techno-economic optimisation framework for decentralised solar-powered cold storage in interconnected open-air meat markets. Using five meat markets in Abuja, Nigeria, the model combines field-derived cooling demand, solar photovoltaic generation, refrigeration, battery storage, phase change material thermal storage, and directed inter-market meat flows. A key feature is that pre-chilled meat is represented as a carrier of product-embodied cooling credit, while phase change material storage remains a stationary cold-side storage component at each market. This allows cooling supplied at one market to reduce the sensible cooling load required at another market. Results show that shifting part of the storage function from battery storage to phase change material thermal storage reduces battery capacity by approximately 67\% and lowers total system cost by up to 15\% compared with battery-only systems. Allowing inter-market cooling-credit exchange further reduces total system cost by 8\% and aggregate phase change material storage capacity by 35\%, mainly by reallocating refrigeration and storage requirements across connected markets. The findings show that product flows can change where cooling services are required, allowing refrigeration and storage capacity to be coordinated across connected sites. Accounting for the product-mediated redistribution of cooling demand extends decentralised energy-system planning beyond isolated demand nodes and may inform cluster-level cooling infrastructure design in other infrastructure-constrained food networks.

\end{abstract}


\begin{highlights}
\item A techno-economic framework for decentralised cold storage in informal markets.
\item Pre-chilled meat is modelled as product-embodied cooling credit between markets.
\item Thermal storage using phase change materials reduces battery dependence and total system cost.
\item Inter-market cooling-credit exchange lowers aggregate storage investment.
\item Food-flow coordination enables cluster-level cold-chain design in weak-grid settings.
\end{highlights}


\maketitle

\section{Introduction}

Nigeria, with over 200 million people, is the most populous country in Africa. Agriculture forms the backbone of its economy, accounting for over 24\% of total gross domestic product (GDP) in 2020 and employing about two-thirds of the population \cite{onwude2023bottlenecks}. Nevertheless, about 40\% of Nigerian households are still food-insecure, where post-harvest losses (PHL) are commonly seen as a significant factor contributing to the gap between food production and food security \cite{adeagbo2026understanding}. Nigeria loses approximately 20-50\% of agricultural output annually \cite{edo2026RoleColdChain}, with associated income losses \cite{onwude2023bottlenecks}, wastage of resources such as labour, land, water, and fertiliser \cite{wani2024new}, greenhouse gas emissions \cite{seth2020nigeria}, and diminished overall food system resilience\cite{bajvzelj2020role}. Within the agricultural sector, livestock production plays a crucial role in supporting food security and nutrition, contributing approximately one-third of Nigeria’s agricultural output \cite{kwaghe2016proper}. In Nigeria, meat is valued not only as a nutrient-dense food, providing protein, vitamins, and minerals, but also as a culturally significant component of everyday diets \cite{smith2022modeling, adenuga2023nigerian}. Meat post-harvest loss in Nigeria is a significant but under-quantified problem. 

Across Nigeria, perishable foods including meat, are predominantly traded through open-air markets, also referred to as informal or traditional markets. These markets consist of networks of small-scale traders and producers who sell agricultural food products, non-food commodities, and related services \cite{dewar2018urban}. They account for more than 80\% of food sales \cite{grace2014taking} and remain central to rural–urban food distribution \cite{hannah2022persistence}. These markets are not isolated retail sites, but interconnected nodes within a wider meat distribution system \cite{valerio2020NetworkAnalysisRegional,akerele2024marketing}, linking primary, secondary, and terminal markets \cite{y.m2018AnalysisChannelStructure,mafimisebi2013FundamentalsCattleMarketing} through daily trade and redistribution routes \cite{valerio2020NetworkAnalysisRegional, akerele2024marketing}. However, cold storage solutions for meat preservation are largely absent in open-air meat markets \cite{adenuga2023nigerian}. As a result, many vendors face daily losses when meat is not fully sold by the end of the day. The price of unsold meat often drops sharply on the second day, particularly when it has been exposed in the open and its fresh red colour has turned dull brown. In the worst cases, the meat spoils and must be discarded.

These losses point to the need for affordable preservation options at the market level. Previous studies have discussed the potential for developing refrigeration in open-air meat markets \cite{odeyemi2021SolarPoweredColdStorage, onaolapo2025effectiveness}. However, the adoption of such systems in infrastructure-constrained settings remains limited by unreliable electricity supply and high investment costs \cite{ali2026SolarDrivenRefrigeration, unep2022SustainableFoodCold}. Refrigeration requires a reliable electricity supply, yet many markets are not connected to grid power. Even where grid connections exist, electricity supply is often unstable, making it difficult to operate conventional cold storage continuously \cite{baurzhan2016off, azimoh2017declining}. Against this background, solar-powered refrigeration has been examined as one viable option for developing decentralised cold-chain systems in weak-grid and off-grid settings. Africa has 60\% of the world’s best solar resources, yet accounts for only 1\% of installed solar PV capacity \cite{iea2022africa}. Nigeria also has substantial solar energy potential, receiving approximately $4.85 \times 10^{12}$ kWh of solar energy per day, equivalent to the energy produced from about 1.1 million tonnes of oil per day \cite{agboSolarEnergyPanacea2021}. This resource potential creates an important opportunity to address post-harvest losses in meat markets through decentralised cooling systems. The other major bottleneck is economic. Open-air meat markets are weakly integrated into formal food-safety, infrastructure, and cold-chain planning \cite{giroux2021informal}. As a result, refrigeration provision in these settings is less likely to depend on fully centralised infrastructure, and may instead require decentralised or market-level solutions \cite{hansen2015review}. Many small-scale traders are unable to afford the high upfront costs \cite{onwude2023bottlenecks, polandinstituteofinternationalbusinessandeconomicsdepartmentoflogisticspoznanuniversityofeconomicsandbusinesspoznanpoland2024ColdSupplyChain}. Although PV costs have decreased \cite{chen2025quantifying}, a major economic and environmental constraint for stand-alone solar cooling systems is the size and cost of battery backup \cite{amjad2023decentralized, amini2025life}. This financial constraint limits the practical adoption of cold-chain technologies, even where solar resources are technically sufficient. Therefore, while solar-powered refrigeration offers a promising infrastructural solution for weak-grid and off-grid meat markets, its feasibility also depends on whether the system can be made economically accessible to small traders. Thermal energy storage has been found helpful for aligning intermittent solar generation with refrigeration demand by storing cooling produced during periods of available electricity and releasing it when cooling is required \cite{he2021technologies}. In refrigeration applications, phase change materials (PCMs) are particularly relevant because they can store latent heat over a narrow temperature range, making them suitable for maintaining chilled conditions \cite{alva2017thermal}. Compared with conventional sensible thermal storage, latent thermal storage can provide the same cooling capacity with a smaller quantity of storage material because energy is stored through the phase-change process \cite{yin2021hydrates}. This higher storage density makes PCMs particularly relevant for decentralised refrigeration systems, where reducing storage volume, equipment size, capital cost, and auxiliary energy consumption is important for practical adoption. Previous studies have examined the use of PCMs in solar-powered refrigeration \cite{guo2025review}, refrigeration-integrated cold thermal storage \cite{selvnes2021review}, and cold-chain transport containers \cite{tong2021phase}, showing their potential as a cold-side buffer that can reduce reliance on battery.

Existing economic studies of cold-chain development in Nigeria have mainly examined refrigeration as a local infrastructure intervention, asking whether access to cold storage can improve market outcomes \cite{takeshima2023solar, pokhrel2020TechnoEconomicTradeOffBattery, okoye2018EconomicFeasibilitySolar}. This framing is useful for assessing the value of refrigeration provision, however, it gives limited attention to how cooling demand may change across interconnected meat-distribution networks when meat is redistributed between markets \cite{singh2018ColdChainConfiguration, sun2025InvestigationMultiobjectiveDecisionmaking, zanoni2019EcoefficientColdChain}. This is a limitation because meat markets are connected through daily redistribution flows, so cooling demand is not necessarily independent across sites. If meat is cooled at an origin market before being transported, the destination market may no longer need to provide the same full sensible cooling load. Treating each market as an isolated cooling site may therefore overestimate refrigeration requirements at some locations, underestimate the system-level value of cooling at others, and overlook potential cost reductions from coordinating refrigeration capacity across the network. 

To address this limitation, this research proposes the concept of product-embodied cooling credit. Unlike conventional mobile cold-storage concepts based on dedicated containers, refrigerated vehicles, or transportable thermal storage units, this concept treats chilled meat not only as a food product, but also as a carrier of avoided cooling demand within the distribution network. 

Under this concept, meat cooled at an origin market carries a thermal state that reduces the sensible cooling requirement at the destination market. The cooling provided in one location can therefore influence cooling demand elsewhere in the network. This reframes decentralised refrigeration from a single-market infrastructure intervention into a network-level system design problem. The value of cooling is no longer assessed only by whether refrigeration improves outcomes within an individual market, but also by how cooling capacity can be allocated across markets to reduce repeated cooling loads, lower system costs, and improve the affordability of cold-chain provision. Building on this concept, this study examines how solar-powered refrigeration can be configured across interconnected open-air meat markets, considering both local cooling demand and the cooling credit embodied in chilled meat flows. In this paper, we present a techno-economic optimisation model for decentralised solar-powered refrigeration in interconnected open-air meat markets. The model is applied to five meat markets in Abuja, Nigeria, and considers local cooling demand, inter-market meat flows, solar PV generation, battery storage, and thermal storage. By incorporating product-embodied cooling credit, the model examines how cooling supplied at one market can affect cooling requirements elsewhere in the network, and how this changes the cost and configuration of decentralised cold-chain systems.

\section{Methodology}
This section describes the methodological framework used to evaluate decentralised cooling across open-air meat markets in Nigeria. It covers the system boundary, input data, cooling-demand calculation, scenario design, model formulation, PyPSA implementation, and sensitivity analysis.

\subsection{Model system and boundary}

\begin{figure}[h]
    \centering
    \includegraphics[width=0.8\linewidth]{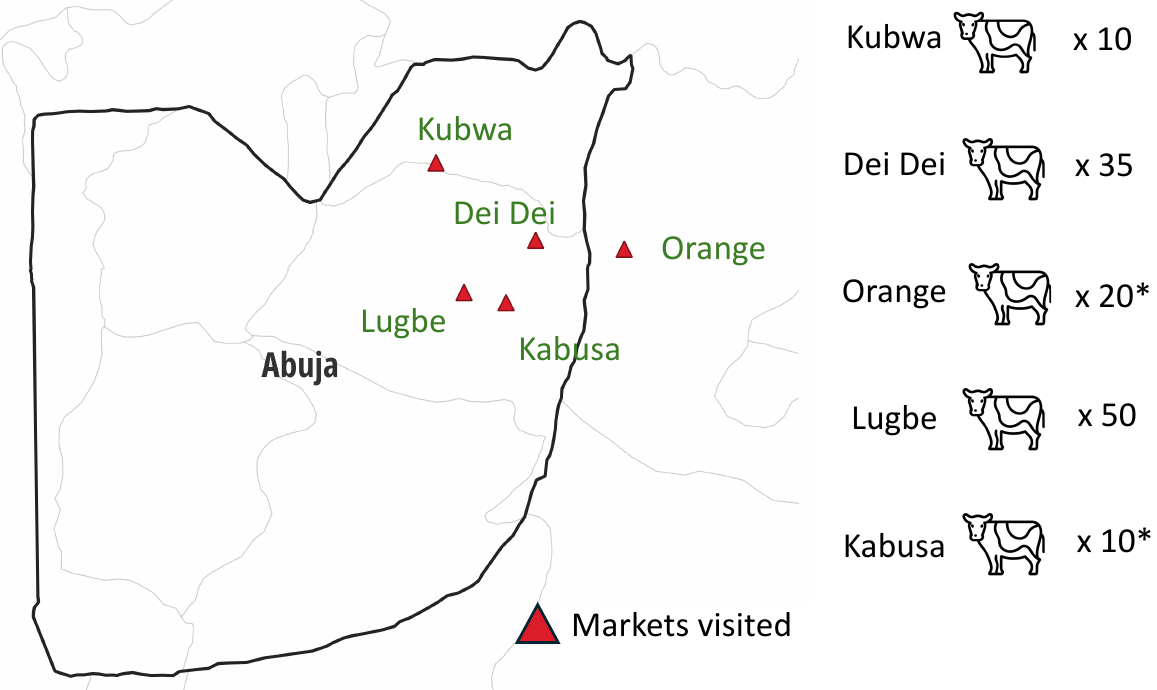}
    \caption{Five open-air meat markets visited in Abuja and their estimated daily cattle throughput provided by the local vendors. Values marked with an asterisk were assumed by the authors based on interviews with meat vendors.}
    \label{fig:market visited}
\end{figure}

\begin{figure}[h]
    \centering
    \includegraphics[width=0.8\linewidth]{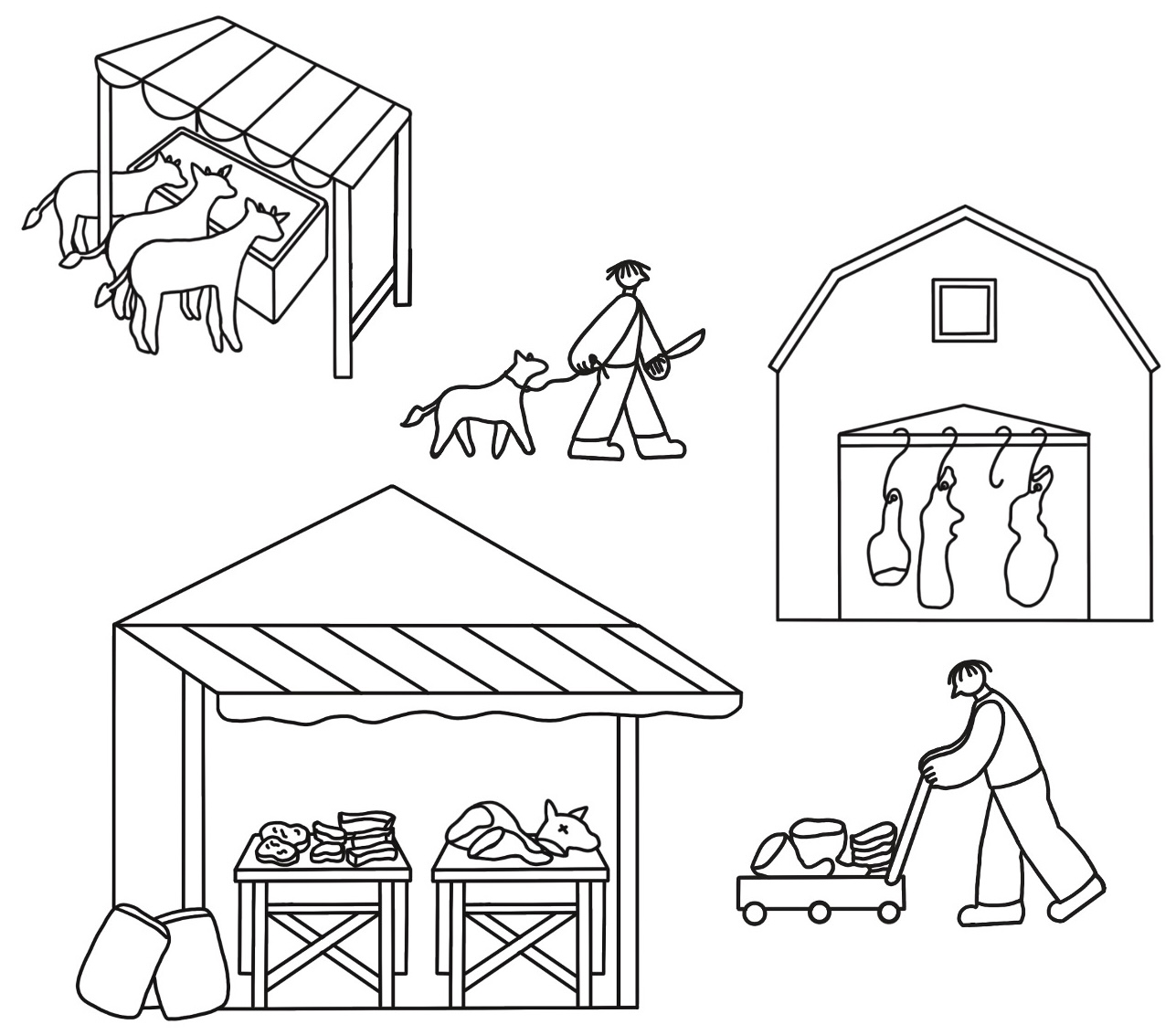}
    \caption{Example of a multifunctional open-air meat market integrating livestock holding, slaughtering, butchering, retail, and short-distance redistribution within a single physical space.}
    \label{fig:Multifunctional market}
\end{figure}

Field investigations were conducted in August 2023 across five major open-air meat markets in Abuja, Nigeria: Kubwa, Dei Dei, Orange, Kabusa, and Lugbe. The selected sites represent variation in market size and location (Figure \ref{fig:market visited}). Field observations showed that the five markets differ in daily cattle throughput and location, but share similar functional structures. Each market combines livestock holding, slaughtering and retail (Figure \ref{fig:Multifunctional market}). Meat products are also transported between markets during trading hours. These observations support the use-study representation of the five markets as an interconnected market-level food-flow network rather than as a set of isolated retail points. A summary of the field investigation evidence supporting this system representation is provided in Supplementary Note S1 and Table S1. Based on this system characterisation, each market is represented as a node with local meat handling activity and cooling demand. Market nodes are connected through directed inter-market transport links that represent existing meat movement between markets. The system boundary includes market-level cooling infrastructure and meat-flow interactions, allowing the model to evaluate decentralised cold-chain configurations under different storage and inter-market coordination assumptions.

\begin{figure}[h]
    \centering
    \includegraphics[width=0.8\linewidth]{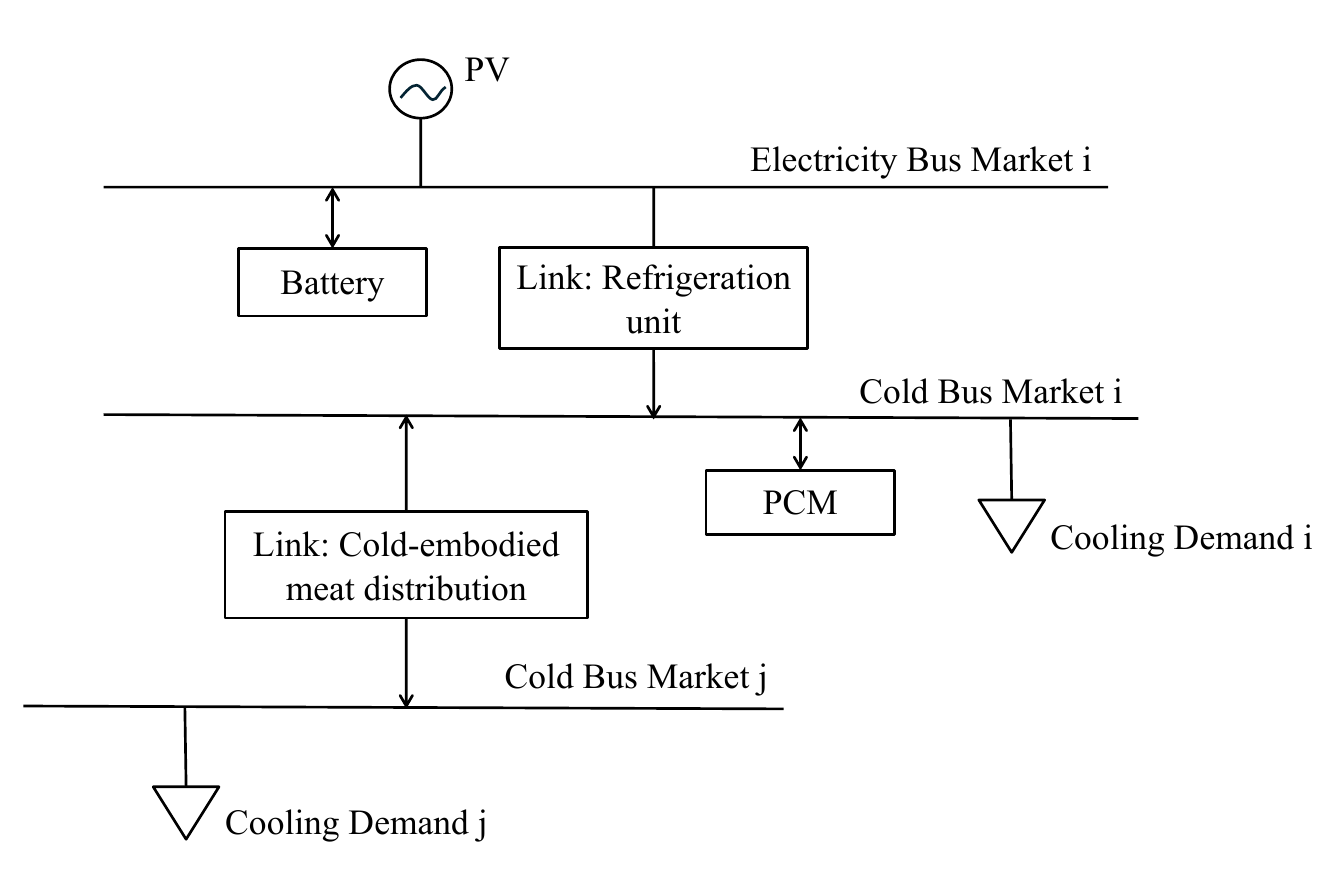}
    \caption{Market-level cold-chain energy system with inter-market transport links. Each market node includes local generator, refrigeration unit and storage, while inter-market transport links connect market cold buses to represent the transfer of cooling-credit embodied in pre-chilled meat.}
    \label{fig:energy flow}
\end{figure}

The system boundary shown in Figure \ref{fig:energy flow} represents the cold-chain energy system for a focal market $i$ and its transport connection to another market $j$. Each market node contains an electricity bus and a cold bus. PV generation supplies the electricity bus, while the vapour-compression refrigeration unit converts electricity into cooling and links the electricity bus to the cold bus. Battery storage is connected to the electricity bus to provide electrical balancing. PCM-based thermal storage is connected to the cold bus and is treated as stationary thermal storage at the market node. Cooling demand is supplied from the cold bus. Inter-market meat movement is represented by directed transport links between market nodes and implemented at the cold-bus level in the energy model. These links capture existing vehicle-based meat movement and the associated distance-based transport cost. In the scenario that allows inter-market cooling-credit accounting, the same links are also used to account for the cooling already supplied to pre-chilled meat before it reaches a destination market. Transport links represent meat-flow interactions only, while cooling infrastructure, including refrigeration units, batteries, and PCM storage, remains located at market nodes. Bidirectional exchange between two markets is represented using separate directed links. Within this boundary, the model evaluates how decentralised cooling infrastructure can be sized and operated under different assumptions about storage and inter-market coordination. The detailed cooling-demand formulation, scenario design, component representation, and optimisation problem are described in the following subsections.

\subsection{Input data and cooling-demand profile}

The main quantitative input for the cooling-demand calculation is the daily cattle throughput at each market. Because such market-level operational data are not available in public datasets, the field-derived values were used to define the relative scale of cooling demand across the five case-study markets. Additional field information on operating hours, existing refrigeration practices, electricity access conditions, and inter-market redistribution patterns was used to support the modelling assumptions. The field observations used to support the market-operation assumptions are summarised in Supplementary Note S1 and Table S1. Daily cattle throughput was converted into meat mass using an average carcass yield of 275 kg per animal \cite{vasko2022opportunities}. The daily sensible cooling requirement at each market was then estimated using Eq. \ref{eq:thermal_load} \cite{lienhard2024heat}.

\begin{equation}
E_{m,d}^{cool} = M_{m,d}c_p\Delta T
\label{eq:thermal_load}
\end{equation}
\noindent where $E_{m,d}^{cool}$ is the daily sensible cooling energy required at market $m$, $M_{m,d}$ the daily meat mass, $c_p$ was set to $8.72 \times 10^{-4}$ kWh/kgK (equivalent to 3.14 kJ/kgK \cite{tavman2007apparent}). $\Delta T$ is the temperature difference. The initial temperature was set to $40^\circ$C, representing an upper-bound ambient condition for freshly handled meat in open-air markets, while the target storage temperature was set to $4^\circ$C\cite{cano1991manual,efsa2014scientific}.

The daily cooling energy was converted into hourly refrigeration demand profiles using operating schedules observed during fieldwork. Two operating regimes were defined: a daytime regime from 04:00 to 19:00, representing slaughtering and active trading hours with higher cooling demand, and a night-time regime from 19:00 to 04:00, representing residual refrigeration of unsold meat. Within each regime, hourly load fractions were generated as non-negative normalised values, preserving the total daily cooling energy while introducing intra-day variability. The resulting hourly refrigeration demand profiles used in the optimisation are provided in Supplementary Note S2 and Figure S1. The cooling-demand calculation assumes uniform product temperature during cooling, constant beef specific heat capacity, negligible microbial heat generation, and no additional external heat gains during the initial cooling process. These assumptions focus the model on the dominant sensible cooling requirement associated with reducing meat from the initial handling temperature to the target refrigerated temperature.

\subsection{Method overview and scenario design}

An hourly techno-economic optimisation model was developed to compare alternative decentralised cold-chain configurations for the five-market network. The model determines the least-cost capacity and operation of the energy-system components defined in Section 2.1 under three scenarios, denoted as S1, S2, and S3. The scenarios are designed to separate the cost effects of battery-based storage, stationary PCM thermal storage, and product-embodied cooling-credit accounting.
\begin{itemize}
    \item S1 (battery-based decentralised systems): each market operates independently, with solar PV and battery storage used to supply refrigeration demand and maintain continuous cooling. No PCM thermal storage or inter-market cooling-credit accounting is included.
    \item S2 (PCM-supported decentralised systems): Stationary PCM-based thermal storage is introduced at each market to provide local thermal buffering and reduce part of the battery storage requirement. Markets remain independent, and transported meat is not credited with prior cooling.
    \item S3 (product-embodied cooling-credit exchange): Building on S2, the model accounts for pre-chilled meat transported between markets through existing trade routes. The cooling already supplied at the origin market is counted as avoided sensible cooling demand at the destination market. Inter-market chilled-meat flows and their associated cooling credits are optimised jointly with generation, refrigeration, battery, and PCM capacities.
\end{itemize}
Across all scenarios, system performance is evaluated using annualised total system cost as the primary metric. Cost components associated with PV generation, refrigeration, battery storage, and PCM thermal storage are analysed separately to identify the role of each technology in the least-cost configuration. 

\subsection{Model formulation and implementation in PyPSA}

An hourly linear techno-economic optimisation model was formulated and implemented using the open-source Python for Power System Analysis (PyPSA) framework \cite{PyPSA}. The model represents decentralised cold-chain systems at the market scale and optimises both installed capacities and hourly operation. Table \ref{tab:tech_costs} summarises the different investment represented in the model. Each market is represented using two energy carriers, electricity and cold, as described in Section 2.1. Additional details of the optimisation structure, including the objective function, electricity balance, cold-energy balance, storage dynamics, and inter-market cold-energy transfer formulation, are provided in Supplementary Note S3.

\subsubsection{PV generator}

The optimisation is performed at hourly resolution over the full year 2019. Solar resource data for Abuja (9.0643°N, 7.4893°E) are obtained from Renewables.ninja \cite{pfenninger2016long}. Because the five markets are located within the same urban area, identical solar availability profiles are assumed for all markets. PV generation is represented using normalised hourly availability factors implemented through the PyPSA parameter $p\_max\_pu$, while each market is allowed to optimise its own installed PV capacity. 

\subsubsection{Storage}

Two storage technologies are represented in the model: battery storage on the electricity carrier and PCM-based thermal storage on the cold carrier. Battery storage is used to shift PV electricity across hours and support refrigeration operation. PCM storage is implemented as stationary thermal storage connected to the cold carrier. It stores cooling produced by the refrigeration unit and releases it later to meet cooling demand. PCM therefore provides local thermal buffering at the market node and can reduce the required battery capacity by shifting part of the storage function from the electricity side to the cold side. PCM storage is enabled in S2 and S3, while S1 relies only on battery storage for temporal balancing. PCM is not transported between markets and is not used as a mobile storage medium. Both storage technologies are implemented as extendable components, while investment capacity and hourly operation are optimised simultaneously within the annual cost-minimisation problem.

\subsubsection{Refrigeration unit}

Refrigeration is represented as a conversion link between the electricity bus and the cold bus. The link consumes electricity and produces cooling according to a fixed coefficient of performance (COP). Installed refrigeration capacity is optimised for each market, and the refrigeration unit must provide sufficient cooling, together with any available storage discharge and cooling credits, to satisfy hourly cooling demand. This representation captures the main energy conversion process required for chilled meat preservation while keeping the model linear.

\subsubsection{Transport}

In S3, inter-market transport is represented using directed PyPSA Links between the cold buses of different market nodes. For each ordered pair of markets $i$ and $j$, where $i\neq j$, a direct link is added from the cold bus of market $i$ to the cold bus of market $j$. The link dispatch represents the product-embodied cooling credit associated with chilled-meat movement. It is not modelled as a refrigerated vehicle, PCM-equipped transport unit, or mobile storage device. Instead, it represents existing diesel-vehicle distribution routes between markets. Transport cost is implemented as the marginal cost of each directed inter-market link. The optimisation therefore penalises each unit of product-embodied cooling credit moved from market $i$ to market $j$ according to a distance-based cost $c_{ij}$. This cost enters the objective function through the link dispatch variable, rather than as a fixed investment cost for transport vehicles or mobile storage equipment. The marginal cost is defined as Eq. \ref{eq:transport}.

\begin{equation}
    c_{ij}= \frac{\alpha _{fuel}d_{ij} }{E_{c}^{cool}}
\label{eq:transport}
\end{equation}
where $c_{ij}$ is the marginal transport cost per unit of product-embodied cooling credit from market $i$ to market $j$, $d_{ij}$ is the haversine distance between the two markets, $\alpha _{fuel}$ is the fuel-based transport cost (0.063 \$/km is used \cite{NBS2019diesel}). $E_{c}^{cool}$ is the cooling credit embodied in a container carrying $m_c$ kg of the pre-chilled meat. It is estimated using the same sensible-cooling calculation as Eq. \ref{eq:thermal_load}, with $m_c$ replacing the daily meat mass $M_{m,d}$. The container mass $m_c$ is treated as a representative transport unit for converting distance-based transport cost into a cost per unit of cooling credit.

The link capacity is modelled as extendable, allowing the optimisation to determine the amount of inter-market chilled-meat movement that is economically justified. Bidirectional exchange between two markets is represented using separate directed links. In the base case, the product-embodied cooling credit is assumed to be fully retained during inter-market movement, providing an upper-bound estimate of the potential system-cost benefit. No additional electricity consumption for transport is modelled.

\begin{table}[h]
\centering
\caption{Technology capital cost parameters used in the optimisation model.}
\label{tab:tech_costs}
\setlength{\tabcolsep}{8pt}
\renewcommand{\arraystretch}{1.35}
\begin{tabular}{lcccc}
\toprule
\textbf{Component}
  & \textbf{Unit capex}
  & \textbf{Lifetime}
  & \textbf{Efficiency}
  & \textbf{Source} \\
\midrule
Solar PV
  & \$580\,/kW$_\text{el}$
  & 25\,yr
  & 1
  & \cite{takeshima2023solar} \\
 
Battery storage
  & \$168.10\,/kWh$_\text{el}$
  & 15\,yr
  & 1
  & \cite{brown2018synergies} \\
 
PCM thermal storage
  & \$19.01\,/kWh$_\text{cold}$
  & 20\,yr
  & 0.8
  & \cite{tong2021phase} \\
 
Refrigeration unit
  & \$164\,/kW$_\text{cold}$
  & 15\,yr
  & 4.83
  & \cite{razmi2020thermoeconomic} \\
\bottomrule
\end{tabular}
\end{table}

\subsection{Sensitivity analysis of thermal storage parameters}
A sensitivity analysis was conducted to evaluate the influence of PCM storage parameters on system cost and technology allocation. Two parameters were varied: PCM capital cost and discharge efficiency. PCM capital cost was varied from \$0.75 to \$168.10 per kWh in \$1 increments, covering reported cost ranges and extending to the cost level of battery storage. Discharge efficiency, defined as the fraction of stored cold energy recoverable to meet demand, was varied from 0.1 to 1.0 in increments of 0.05. For each parameter combination, the optimisation model was solved for all five markets while holding all other inputs constant. No structural changes were introduced relative to the baseline formulation. The resulting total system cost, installed storage capacities, refrigeration capacity, and refrigeration runtime were recorded for comparison.

\section{Results}

\subsection{Cost comparison between battery-based and PCM-based cold storage}
\begin{figure}[h]
    \centering
    \includegraphics[width=\linewidth]{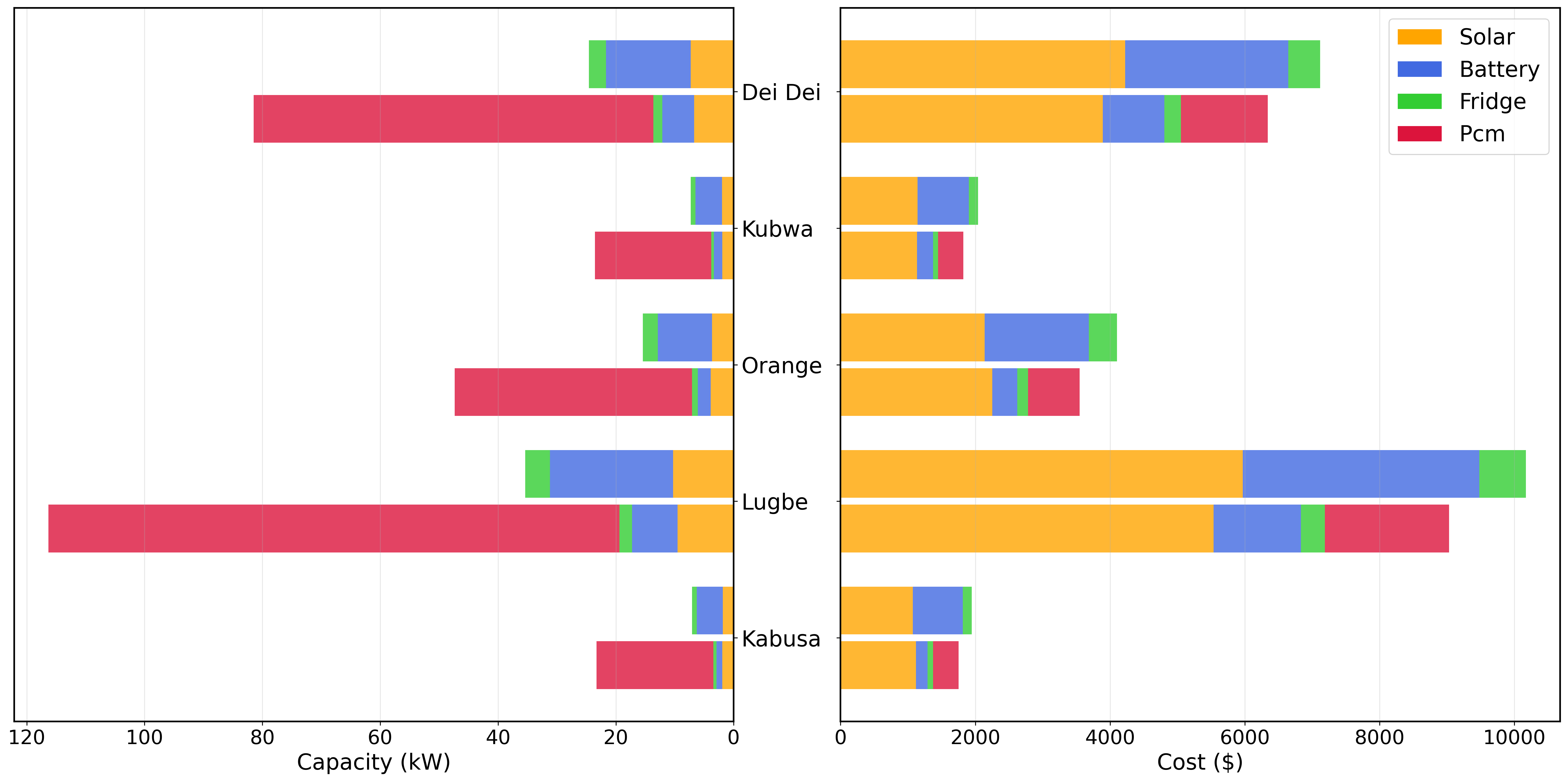}
    \caption{Capacity and cost comparison between S1 and S2}
    \label{fig:dual stacked bar chart}
\end{figure}

\begin{figure}[h]
    \centering
    \begin{subfigure}[b]{0.48\linewidth}
        \centering
        \includegraphics[width=\linewidth]
        {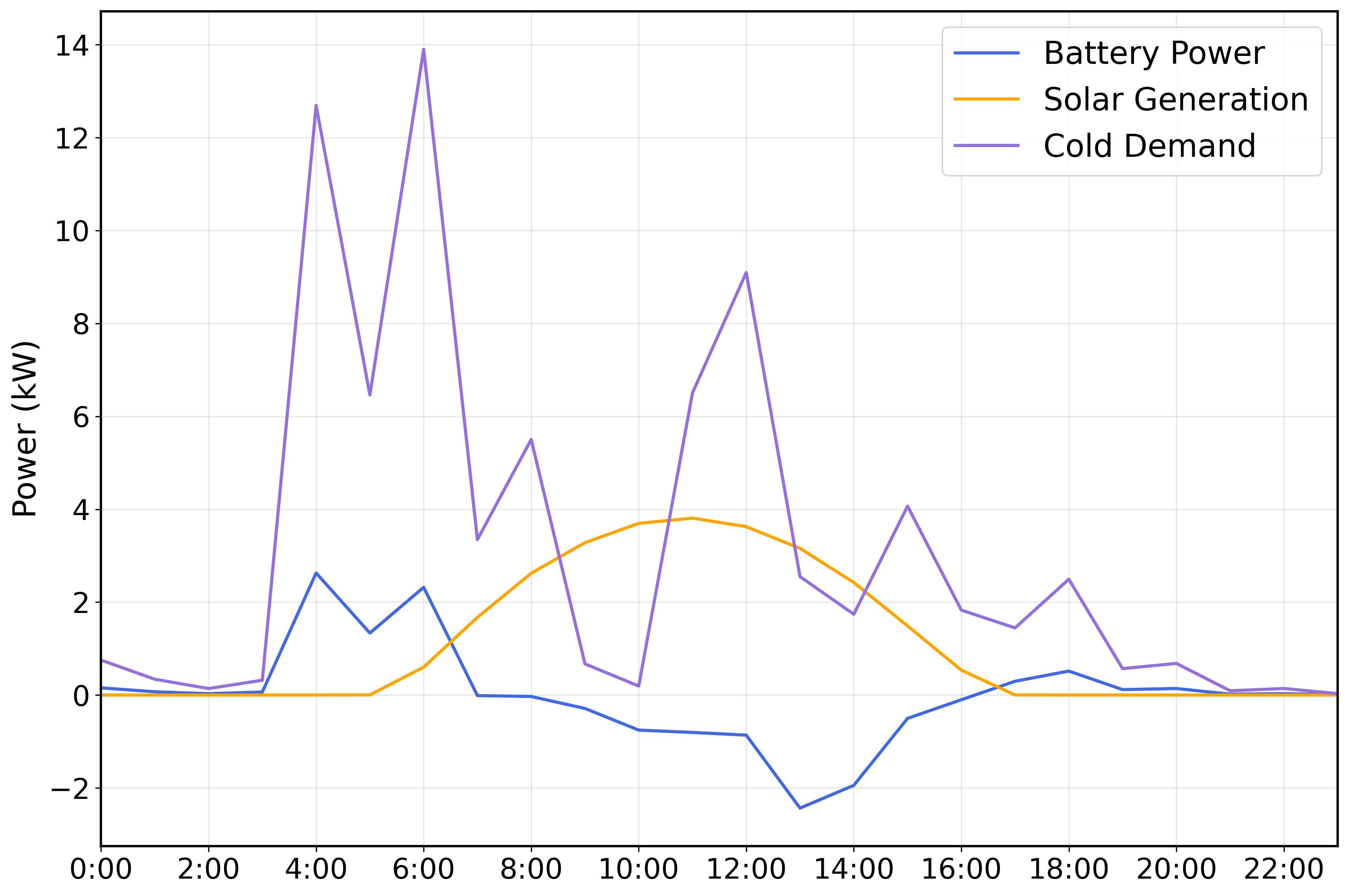}
        \caption{Electricity storage scenario (S1)}
        \label{fig:cold_demand_supply_a}
    \end{subfigure}
    \hfill
    \begin{subfigure}[b]{0.48\linewidth}
        \centering
        \includegraphics[width=\linewidth]
        {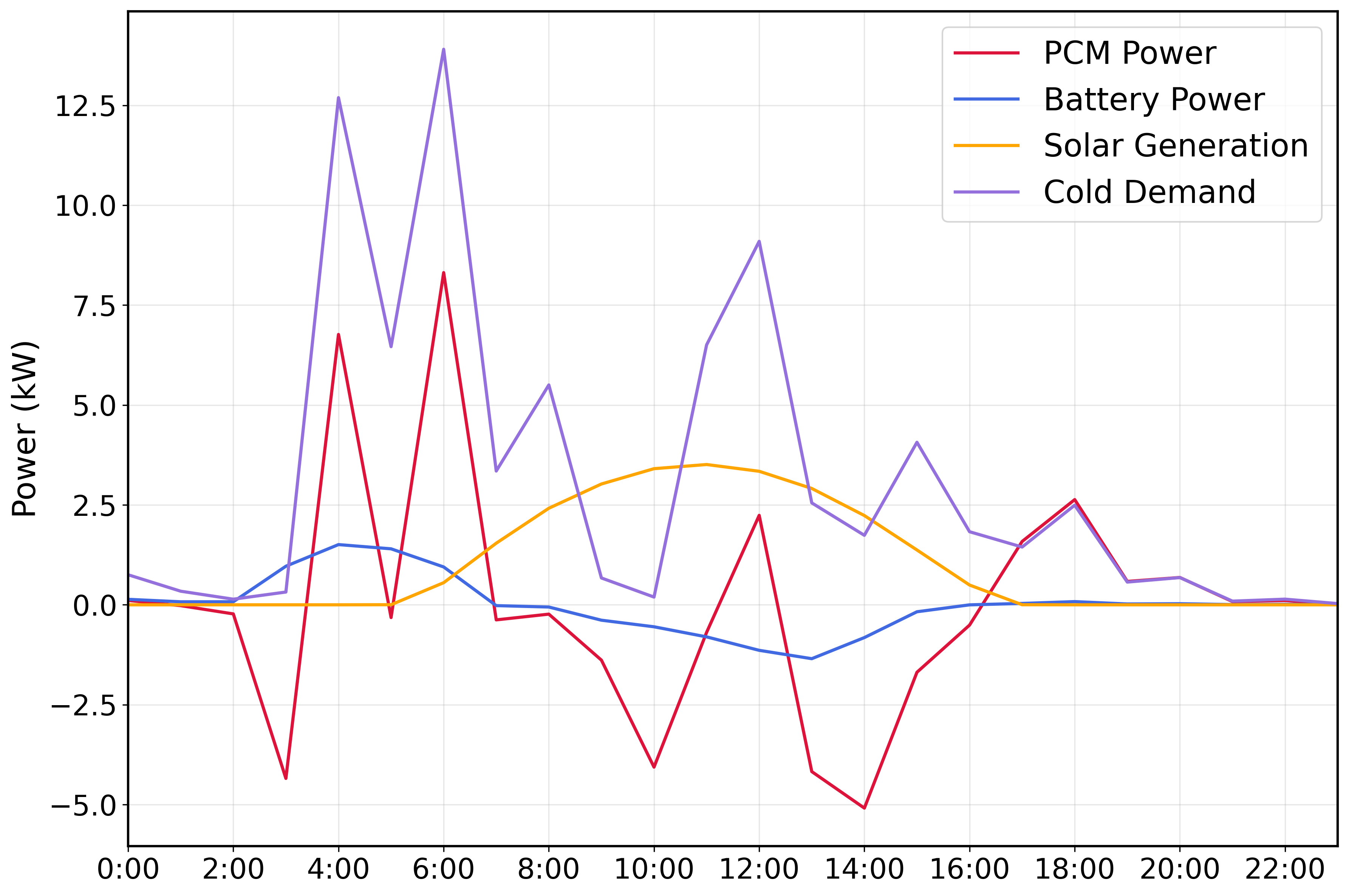}
        \caption{Thermal storage scenario (S2)}
        \label{fig:cold_demand_supply_b}
    \end{subfigure}
    \caption{Cold demand and supply alignment comparison}
    \label{fig:cold_demand_supply_combined}
\end{figure}

Compared with S1, S2 reduces total annualised system cost across all five markets. The maximum cost reduction reaches 15\%, and the average reduction across markets is 11\%. This reduction is mainly associated with the replacement of part of the electrical storage function by thermal storage on the cold side of the system. The total cooling demand remains unchanged between S1 and S2, but the introduction of PCM changes how cooling is produced and stored over time. In S2, part of the cooling generated during periods of solar availability is stored as thermal energy and released later to meet demand, reducing the need for battery storage during non-solar hours. Figure \ref{fig:dual stacked bar chart} shows the optimised capacities under both scenarios. Averaged across the five markets, battery capacity decreases by approximately 67\% when PCM storage is introduced. This is accompanied by a modest increase in PV capacity of around 4\%, which reflects the additional daytime generation used to charge thermal storage. Refrigeration capacity also decreases by roughly 50\%, indicating that the system relies less on high electrical storage capacity to support cooling outside solar-generation periods. Although S2 introduces an additional PCM investment and slightly larger PV capacity, these increases are more than offset by the reduction in battery capacity, leading to lower total system cost. Hourly operation patterns in Figure \ref{fig:cold_demand_supply_combined} illustrate the underlying mechanism. In S1, early morning and evening cooling demand is supplied primarily through battery discharge because solar generation and refrigeration demand occur at different times of the day. In S2, solar electricity is used during the daytime to produce cooling and charge PCM storage which is later discharged to meet evening cooling demand. Battery use is therefore reduced and mainly retained for short-term balancing rather than serving as the main storage medium.

\subsection{Inter-market cold energy flows and system cost outcomes}
Enabling inter-market product-embodied cooling-credit exchange in S3 further reduces total system cost compared with the isolated PCM-supported configuration in S2. At the system level, total annualised cost decreases by 8\%,  while aggregate PCM storage capacity decreases by 35\% (Table \ref{tab:cold_energy_transfer_impacts}). Total PV capacity remains almost unchanged. This indicates that the main benefit of inter-market exchange is not the expansion of generation capacity, but the reduction and redistribution of storage requirements across the market network. The cost effects vary across individual markets. Kabusa and Orange show the largest market-level cost reductions, both decreasing by 4\%. Lugbe and Dei Dei show smaller reductions, while Kubwa experiences a 3\% increase in cost. This increase is associated with higher PV and refrigeration capacities at Kubwa, despite a reduction in its PCM capacity. The result suggests that Kubwa takes on a stronger exporter role in the optimised network, producing and transferring part of the cooling credit used by other markets. Table \ref{tab:exchange_index} shows that both imports and exports of cooling credit occur across the market network. These flows allow some cooling demand to be met through pre-chilled meat transferred from other markets, rather than requiring each market to size its own PCM storage independently for local demand. The optimisation therefore shifts part of the storage burden from individual market-level PCM units to inter-market cooling-credit flows. This explains why aggregate PCM capacity falls while PV capacity remains nearly constant.

\begin{table}[h]
\centering
\caption{Relative changes (\%) in system cost and capacity decisions when enabling inter-market cooling-credit transfer via chilled meat, compared with S2. Negative values indicate reductions.}
\label{tab:cold_energy_transfer_impacts}
\begin{tabular}{lccccc}
\toprule
\makecell{Market}
& \makecell{Total\\Cost}
& \makecell{PV\\Capacity}
& \makecell{Battery\\Capacity}
& \makecell{PCM\\Capacity}
& \makecell{Fridge\\Capacity} \\
\midrule
Kabusa  & -4 & -3 & 15 & -16 & -9  \\
Lugbe   & -1 & 1  & -12 & -2 & 6  \\
Orange  & -4 & -3 & 6 & -12 & -6 \\
Kubwa   & 3  & 6  & 3 & -11 & 11  \\
Dei Dei & -2 & -2 & -18 & 6  & 8  \\
\midrule
Total   & -8 & 0  & -6 & -35 & 11  \\
\bottomrule
\end{tabular}
\end{table}

\begin{table}[htbp]
\centering
\caption{Relative export and import exchange indices across markets. The export and import indices represent each market's relative contribution to total exported and imported cooling-credit exchange, respectively. Positive values indicate a stronger exporter role, while negative values indicate a stronger importer role.}
\label{tab:exchange_index}
\begin{tabular}{lccc}
\toprule
Market & Export share & Import share & Net orientation\\
\midrule
Kabusa  & 21 & 35 & -14 \\
Lugbe   & 9  & 6  & 3 \\
Orange  & 23 & 35 & -12 \\
Kubwa   & 31 & 11 & 20 \\
Dei Dei & 15 & 12 & 3 \\
\bottomrule
\end{tabular}
\end{table}

\subsection{Sensitivity of system cost and storage allocation to PCM parameters}

\begin{figure}[h]
    \centering
    \includegraphics[width=\linewidth]{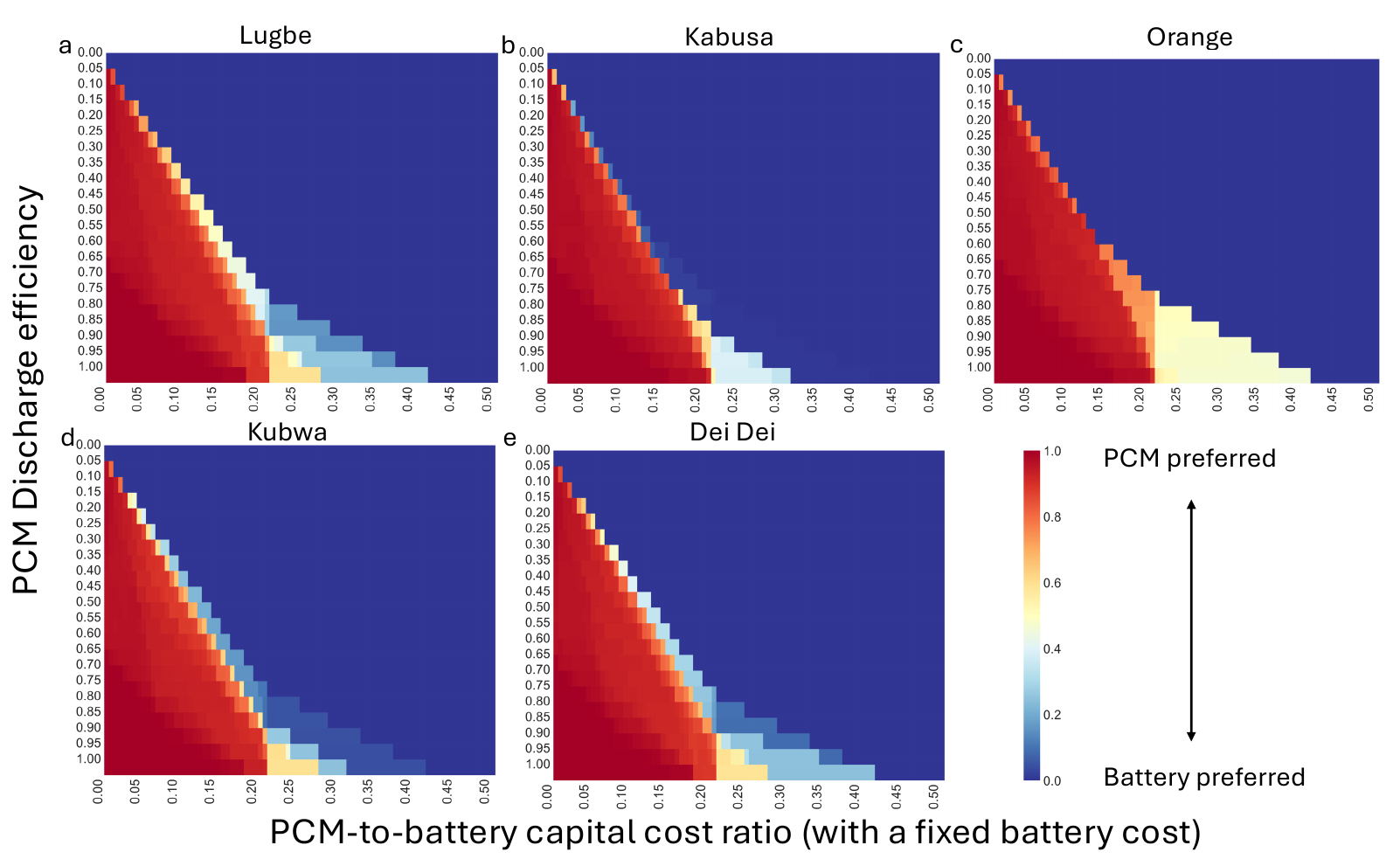}
    \caption{Effect of PCM efficiency and price on the choice between PCM and battery storage.}
    \label{fig:pcm sensitivity}
\end{figure}

The sensitivity analysis shows that PCM adoption depends jointly on capital cost and discharge efficiency. Under S2, PCM is selected when its capital cost is sufficiently low relative to battery storage and when its discharge efficiency is high enough to recover a useful share of stored cold energy. As PCM cost increases or discharge efficiency decreases, the optimisation shifts progressively toward battery-dominated storage solutions. Figure \ref{fig:pcm sensitivity} shows a clear transition frontier between PCM-preferred and battery-preferred solutions. The position of this frontier differs across markets. Lugbe, which has the largest cooling demand, shows the widest PCM-preferred region, meaning that PCM remains cost-effective over a broader range of cost and efficiency assumptions. Orange and Dei Dei show intermediate behaviour, while Kabusa and Kubwa favour PCM only under more restrictive conditions. This pattern indicates that larger cooling loads make thermal storage more economically attractive, because the fixed system benefits of shifting cooling through PCM can be used more intensively.

\begin{figure}[h]
    \centering
    \includegraphics[width=0.8\linewidth]{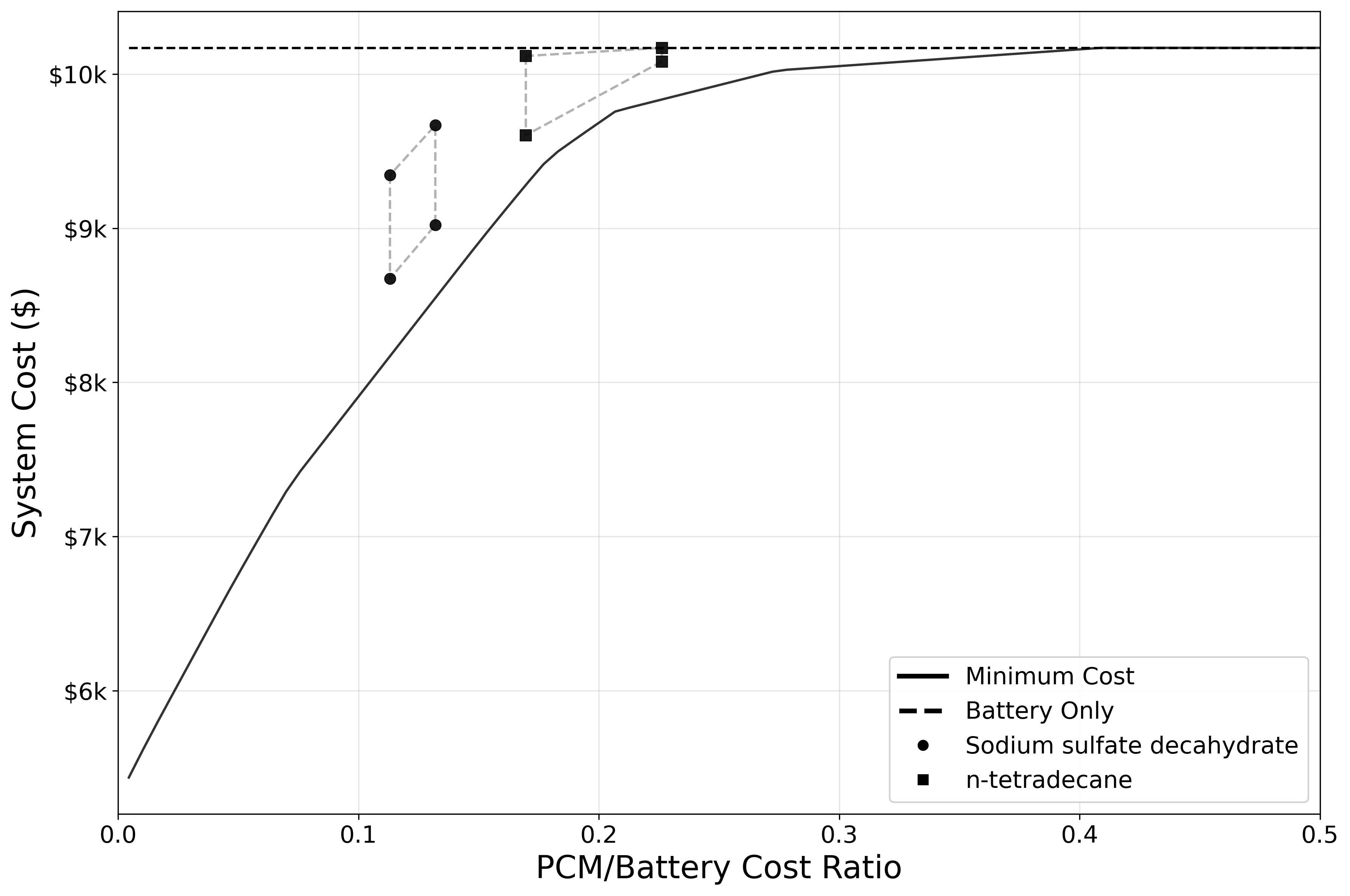}
    \caption{Benchmarking real PCM candidates against the model-derived minimum-cost envelope in Lugbe. The solid curve shows the minimum system cost across the PCM cost--efficiency parameter sweep, while the dashed line indicates the battery-only benchmark. Markers represent reported cost and discharge efficiency ranges for sodium sulfate decahydrate and n-tetradecane (C14).}
\label{fig:pcm_benchmark_lugbe}
\end{figure}

To place the model-derived frontier in relation to realistic PCM options, two representative materials are compared in Table \ref{tab:pcm_properties_cost} and Figure \ref{fig:pcm_benchmark_lugbe}. Lugbe is used for this comparison because it has the largest cooling demand and the widest PCM-preferred region in the sensitivity results. Both sodium sulfate decahydrate and n-tetradecane fall within a reported discharge-efficiency range of 0.7–0.9, but they differ in relative capital cost. Sodium sulfate decahydrate corresponds to a lower PCM-to-battery cost ratio of approximately 0.11–0.13, while n-tetradecane corresponds to a higher ratio of approximately 0.17–0.23. Figure \ref{fig:pcm_benchmark_lugbe} shows that sodium sulfate decahydrate lies closer to the minimum-cost envelope, whereas n-tetradecane is positioned nearer to the battery-only benchmark as the PCM cost ratio increases. This indicates that, within the considered efficiency range, relative capital cost has a stronger influence on PCM competitiveness than moderate differences in discharge efficiency.

\begin{table}[h]
\centering
\caption{Properties and cost parameters of selected PCMs \cite{tong2021phase}}
\label{tab:pcm_properties_cost}
\small
\setlength{\tabcolsep}{4pt}
\begin{tabular}{lcccccc}
\toprule
\makecell{Type} 
& \makecell{Phase Change\\Temperature ($^\circ$C)} 
& \makecell{Density\\(kg/m$^3$)} 
& \makecell{Latent Heat\\(kJ/kg)} 
& \makecell{Discharge\\Efficiency} 
& \makecell{Unit Price\\(\$/ton)} 
& \makecell{Capital Cost\\(\$/kWh)} \\
\midrule
\makecell{n-tetradecane\\(C14)} 
& 5   & 767  & 250 & 0.7--0.9 & 1980--2640 & 28.51--38.02 \\
\makecell{Sodium sulfate\\decahydrate} 
& 7.5 & 1350 & 150 & 0.7--0.9 & 792--924   & 19.01--22.18 \\
\bottomrule
\end{tabular}
\end{table}

\begin{figure}[h]
    \centering
    \includegraphics[width=\linewidth]{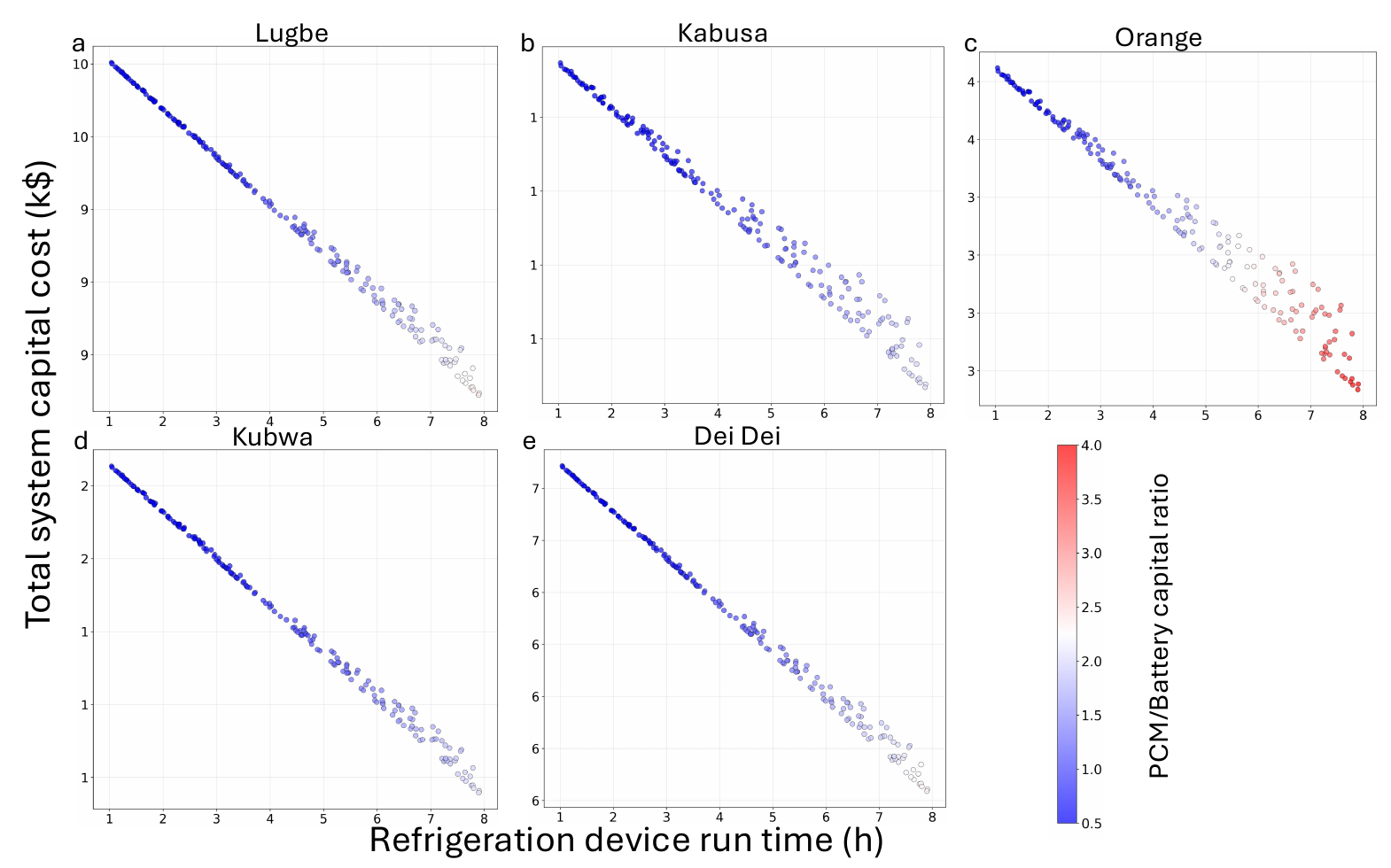}
    \caption{Effect of refrigeration-device runtime and PCM capacity on total system cost.}
    \label{fig:refrigeration devices vs pcm capacity}
\end{figure}

The relationship between optimised refrigeration-device runtime and storage allocation is shown in Figure \ref{fig:refrigeration devices vs pcm capacity}. Total system cost decreases as the available charging window increases. Short charging durations (below approximately 3–4 hours) correspond to battery-dominated solutions, as meeting daily cooling demand within a narrow time window requires high instantaneous charging power. In these cases, batteries provide short-timescale flexibility without necessitating oversized refrigeration capacity. As charging time increases, peak power requirements decline, enabling a larger share of storage to be provided by PCM, reflected in rising PCM-to-battery capacity ratios and lower overall system cost.

\section{Discussion}

The results point to two mechanisms that shape the least-cost design of decentralised cold-chain systems. The first is a change in storage form: when PCM is introduced, part of the storage function shifts from electrical storage to cold-side thermal storage, reducing the role of batteries from the main storage medium to a short-term balancing component. The second is a change in network allocation: when pre-chilled meat moves between connected markets, cooling supplied at the origin can reduce the cooling required at the destination. These mechanisms suggest that cold-chain design should not treat each market only as an isolated PV--battery--refrigeration unit. The design problem also concerns where cooling should be produced, the form in which it should be stored, and how existing food-distribution links may affect cooling demand across market nodes.

\subsection{Reconfiguring storage in decentralised cold systems}

PCM changes the storage pathway by allowing cooling produced during solar-available hours to be stored directly on the cold side. This reduces the need to store electricity before refrigeration and leaves batteries mainly responsible for short-term electrical balancing. Batteries retain broader electrical flexibility, whereas PCM provides a dedicated thermal service. This narrower role can be advantageous when demand is predictable and predominantly thermal. The meat-market case has this feature because cooling demand is linked to daily slaughtering, trading, and overnight storage routines. Under these conditions, storing cold directly can be cheaper than storing electricity first and producing cold later, especially when PCM capital cost is much lower than battery storage cost.

This substitution is also conditional. PCM is favoured only when its lower capital cost is not offset by poor discharge performance. If discharge efficiency is low, more cooling must be produced to deliver the same useful cold output. This can increase refrigeration operation, PV generation, or residual battery use. In that case, the cost advantage of PCM becomes weaker. Similarly, if PCM capital cost approaches battery cost, the benefit of using a specialised thermal storage medium becomes smaller because batteries still provide broader operational flexibility. The transition between PCM-preferred and battery-preferred configurations therefore reflects a trade-off between low-cost thermal storage and flexible electrical storage.

The relationship between refrigeration-device runtime and storage allocation provides further explanation. When cooling must be produced within a short operating period, the system needs high instantaneous cooling capacity or fast storage response. Batteries are better suited to this short-timescale balancing role. When refrigeration can operate over a longer period, cooling production can be spread across solar-available hours and stored in PCM. This reduces peak power requirements and allows PCM to provide a larger share of the storage function. PCM is therefore most useful when cooling can be pre-produced over a sufficiently long daytime window, while batteries remain important when operating schedules are tighter or cooling demand is less predictable.

\subsection{Inter-market cooling credit through pre-chilled meat flows}
Product-embodied cooling credit should be distinguished from both conventional sensible-heat storage and PCM-based thermal storage. Conventional sensible-heat storage stores cooling by changing the temperature of a dedicated storage medium, while PCM storage uses a dedicated phase-change material located at the market node. Product-embodied cooling credit adds no dedicated storage medium. Instead, it recognises that meat cooled for preservation has already undergone part of the required sensible cooling process, reducing the cooling requirement when it reaches another market. The product therefore carries an avoided cooling requirement created by prior heat removal. Product-embodied cooling credit is complementary to PCM storage rather than a direct substitute for conventional sensible-heat or PCM-based storage.

The cost benefit of inter-market cooling credit arises because connected-market design relaxes the need for each market to size its cooling and storage assets against local demand alone. In an isolated design, each node must provide its own PV generation, refrigeration capacity, battery storage, and PCM storage to meet its local cooling profile. When pre-chilled meat can move through existing redistribution routes, part of this local storage burden can be shared across the network. Markets with more favourable generation, refrigeration, or demand conditions can supply cooled products to other markets, while receiving markets can reduce their own PCM requirement. This indicates that the main value of the inter-market mechanism lies in reallocating storage and refrigeration services across the network, rather than increasing solar electricity production. 

The heterogeneous market-level results are central to this mechanism. Some markets benefit mainly as importers of cooling credit, while others may take on an exporter role and require additional PV or refrigeration capacity. Market-level effects are uneven. Some markets benefit mainly as importers of cooling credit, while others take on an exporter role and require additional PV or refrigeration capacity. Kubwa, for example, faces higher local investment because it supplies part of the cooling credit used elsewhere in the network. The system optimum can therefore assign different functional roles to individual markets and reduce total network cost without reducing costs at every node.

The use of product-embodied cooling credit is therefore conditional on both physical and institutional factors. It requires short enough transport distances, sufficient retention of cooling during transport, and timing that allows pre-chilled meat to offset demand at the destination. Its implementation would also require coordination among markets, transporters, and storage operators.

\subsection{Broader energy-system and policy implications}
The findings have two implications for decentralised energy-system planning. First, storage should be selected according to the service required. For predictable cooling loads, PCM can reduce dependence on battery storage by storing cold directly, while batteries remain important for short-term electrical balancing. Second, cooling demand need not be treated as fixed at isolated sites. Battery and PCM storage shift electricity or cooling over time within a market, whereas pre-chilled product flows alter where part of the cooling requirement is met across the network. Considering both mechanisms supports cluster-level planning of refrigeration and storage rather than independent site-by-site design.

Several structural conditions represented by the Abuja case are also found elsewhere in Nigeria and across African urban food systems. More than 40 million urban Nigerians have little or no access to electricity for food refrigeration or cooling, while the substantial increase in national cold-chain capacity between 2014 and 2018 was concentrated in Lagos \cite{seth2020nigeria}. Evidence from four Kenyan and fourteen Zambian cities similarly shows that households predominantly obtain food from open-air markets and informal vendors, with only a subset purchasing food from supermarkets \cite{hannah2022persistence}. The combination of informal market structures, limited cooling access, and uneven infrastructure provision is therefore not specific to limited markets.

The transferable element of the study is the planning approach rather than the numerical results obtained for Abuja. The framework is most relevant where multiple markets or collection points are connected by regular product redistribution, cooling infrastructure is costly or unevenly distributed, and prior cooling can reduce demand at the destination.

The wider relevance is linked to the scale of unmet cold-chain demand. In 2017, the lack of effective refrigeration was estimated to cause the loss of 526 million tonnes of food, equivalent to 12\% of global food production, while food cold chains and food loss caused by inadequate refrigeration together accounted for approximately 4\% of global greenhouse-gas emissions \cite{unep2022SustainableFoodCold}. The Abuja results indicate that coordinating product flows can reduce repeated storage provision and lower the cost of renewable-powered cooling, providing one possible pathway for improving access to cold-chain infrastructure in weak-grid and off-grid regions. If wider access to cooling reduces food loss, further benefits may arise from avoiding the energy and other resources embodied in discarded products and the associated greenhouse-gas emissions. 

\subsection{Stakeholder and governance implications}
For vendors and market actors, the main potential benefit is a lower capital burden for decentralised cooling. Reducing dependence on batteries through PCM storage can make shared cold storage more affordable for trader groups, market associations, or small operators. This may reduce spoilage risk, extend selling time, and improve inventory control, particularly in open-air meat markets where unsold meat can lose value quickly without refrigeration.

These benefits may not be distributed evenly across the market network. Markets that supply cooling credit may require additional refrigeration or PV investment, while receiving markets can reduce their own storage requirements. The system-level optimum may therefore differ from the interests of individual actors. Implementation may require tariffs, service fees, shared ownership, or cooperative arrangements to allocate costs and benefits, together with agreed transport, handling, and temperature-monitoring practices. The model does not determine which governance arrangement would be most appropriate, but the results show that such arrangements would affect whether the network-level configuration could be implemented in practice.

\subsection{Limitations and future work}

Several limitations should qualify the findings. The inter-market case should be interpreted as an upper-bound estimate of system-cost benefit because the current model does not explicitly include heat gains during transport, time-dependent temperature change, route constraints, or vehicle energy use. These factors would reduce the amount of cooling credit retained during transport and may limit when and where product-embodied cooling can be used. The cooling-credit mechanism would still be relevant, but its quantitative benefit would likely be smaller under more detailed transport modelling.

The cooling-demand formulation is also simplified. The model focuses on sensible cooling for meat and does not explicitly represent microbial growth, quality decay, product heterogeneity, or the effect of repeated handling. In practice, the value of cooling depends not only on the amount of heat removed, but also on whether the temperature history is sufficient to maintain food safety and product quality. Future models should link thermal-energy accounting with time-temperature quality dynamics so that cooling credit is assessed both thermodynamically and biologically.

The empirical scope is limited to five open-air meat markets in Abuja. Future work should test the framework in larger regional networks and for other products, including fish, dairy, and fresh produce, whose thermal properties, quality constraints, and economic values may alter the role of PCM storage and cooling-credit exchange.

The model does not represent ownership, payment, or benefit-sharing arrangements. Field-based pilots and stakeholder studies are needed to assess whether least-cost network configurations can be implemented where costs and benefits are unevenly distributed.

\section{Conclusion}
This study developed a techno-economic optimisation framework to evaluate decentralised solar-powered cold storage systems in Nigeria’s informal meat markets. By combining field observations, cooling-demand estimation, and energy-system optimisation, the framework examined how PV generation, refrigeration, battery storage, PCM thermal storage, and inter-market food flows jointly shape least-cost cold-chain configurations. A key feature of the model is that chilled meat is represented not only as a food product, but also as a carrier of product-embodied cooling credit within an interconnected market network.

The results of the Abuja use-study show that PCM thermal storage can reduce battery dependence by 67\% and lower total system cost by up to 15\% when PCM capital cost, discharge efficiency, and refrigeration charging time are favourable. Inter-market cooling-credit exchange further reduces aggregate storage requirements by allowing cooling supplied at one market to offset part of the cooling demand elsewhere, leading to an additional 8\% reduction in total system cost. These findings indicate that decentralised cold-chain design should consider both cold-side storage within markets and inter-market cooling-credit exchange through existing food flows.

The proposed framework has practical application prospects for weak-grid and off-grid food systems where conventional cold-chain infrastructure is difficult to deploy. It can support market-level or cluster-level planning by identifying where cooling infrastructure should be located and how existing food-distribution routes may contribute to cooling provision. It also provides a bottom-up pathway for developing sustainable food cold chains through distributed market-level refrigeration, local thermal storage, and existing food-flow connections, without relying solely on full centralised grid expansion. Because markets are represented as connected energy-system nodes, the same framework can be used to assess future links among market clusters and with national or regional grids. Beyond open-air meat markets in Nigeria, the modelling logic could be applied to other perishable food networks, including fish, dairy, fruit, and vegetable supply chains, especially in regions where food logistics are active but energy infrastructure remains limited.

The current study remains a conceptual and techno-economic assessment. Its main limitations are the simplified treatment of inter-market transport, thermal losses, product-quality change, and market governance, as well as the small empirical scope of five Abuja markets. Future work should test the framework in larger and more diverse food-distribution systems, link cooling-credit accounting with time-temperature quality dynamics, and examine ownership and pricing arrangements through field-based pilots. These extensions would help move the framework toward practical decision support for decentralised cold-chain development in infrastructure-constrained food systems.

\printcredits

\bibliographystyle{elsarticle-num}

\bibliography{ade_ref}

\end{document}